\documentclass[11pt]{article}

\usepackage[top=50mm, bottom=50mm, left=50mm, right=50mm]{geometry}
\usepackage{enotez}
\usepackage{amssymb}
\usepackage{hyperref}
\usepackage{amsmath}
\usepackage{amsrefs}
\usepackage{amsthm}
\usepackage{epsfig}
\usepackage{graphicx}
\usepackage{float}
\usepackage{xurl}
\usepackage{subfigure}
\usepackage{multirow}
\usepackage{color}
\usepackage{lineno}
\usepackage{fullpage}
\usepackage[normalem]{ulem} 
\usepackage{xspace}
\usepackage [english]{babel}
\usepackage [autostyle, english = american]{csquotes}
\MakeOuterQuote{"}
\usepackage{titlesec}

\titleformat{\paragraph}
{\normalfont\normalsize\bfseries}{\theparagraph}{1em}{}
\titlespacing*{\paragraph}
{0pt}{3.25ex plus 1ex minus .2ex}{1.5ex plus .2ex}

\makeindex

\begin{document}

\title{Session: End-To-End Encrypted Conversations With Minimal Metadata Leakage}
 
\author{
{Kee Jefferys\footnote{CTO --- Session : kee@getsession.org}} \and {Maxim Shishmarev\footnote{Software Engineer : maxim@getsession.org}}\and {Simon Harman\footnote{Director --- OPTF : simon@optf.ngo}}
}

\maketitle
\thispagestyle{empty}
\setcounter{page}{0}

\begin{abstract}\noindent Session is an open-source, public-key-based secure messaging application which uses a set of decentralised storage servers and an onion routing protocol to send end-to-end encrypted messages with minimal exposure of user metadata. It does this while providing the common features expected of mainstream messaging applications, such as multi-device syncing, offline inboxes, and voice/video calling. 
\end{abstract}

\renewcommand{\abstractname}{Note to Reader}

\begin{abstract}\noindent 
Parts of this document refer to features which are in the process of being implemented or renamed; see the addendum in section \ref{sec:Addendum} for a full list of these items.
\end{abstract}

\section{Introduction}
Over the past 10 years, there has been a significant increase in the usage of instant messengers, with the most widely-used messengers each having amassed over 1 billion users~\cite{Popular_Messengers}. The potential privacy and security shortfalls of many popular messaging applications have been widely discussed~\cite{Smartphone_Messengers}. Most current methods of protecting user data privacy are focused on encrypting the contents of messages, an approach which has been relatively successful. The widespread deployment of end-to-end encryption does increase user privacy; however, it largely fails to address the growing use of metadata by corporate and state-level actors as a method of tracking user activity. 

In the context of private messaging, metadata can include the IP addresses and phone numbers of the participants, the time and quantity of sent messages, and the relationship each account has with other accounts. Increasingly, it is the existence and analysis of this metadata that poses a significant privacy risk to journalists, protesters, and human rights activists~\cite{Journalist_Metadata}.

Session is, in large part, a response to this growing risk: it provides robust metadata protection on top of existing cryptographic protocols which have already been proven to be effective in providing secure communication channels. 

Session reduces metadata collection in three key ways:

Firstly, Session does not require users to provide a phone number, email address, or other similar identifier when registering a new account. Instead, pseudonymous public-private key pairs are the basis of an account’s identity.

Secondly, Session makes it difficult to link IP addresses to accounts or messages sent or received by users, through the use of an onion routing protocol. 

Thirdly, Session does not rely on central servers; a decentralised network of thousands of economically incentivised nodes performs all core messaging functionality. For those services where decentralisation is impractical, like storage of large attachments and hosting of large group chat channels, Session allows users to self-host infrastructure, and rely on built-in encryption and metadata protection to mitigate trust concerns.

\section{Threat Model}

Before explaining Session’s design, it is useful to understand the scope of threats which it is designed to defend against and the protections Session provides to users.

\subsection{In Scope}

\subsubsection{Session Node Operators --- Passive/Active Attacks}

The Session Node network is permissionless, meaning any operator may join the network if they have a sufficient stake. Session's threat model considers a highly resourced, but financially limited attacker who can only control a fraction of the Session Node network. Session Nodes are responsible for storing messages and operating the onion routing system used by Session called onion requests (see section \ref{sec:session-nodes} and \ref{sec:onion-requests}). A dishonest Session Node operator would be able to perform a range of active or passive attacks. 

Possible passive attacks include reading message headers, logging message relay or receipt timestamps, saving the encrypted contents of messages, or recording the sizes of messages. Possible active attacks include failing to relay messages, failing to store messages, providing clients with modified messages, or refusing to respond to requests for messages belonging to specific Account IDs. 

Possible active attacks on the onion requests system include dropping arbitrary requests, modifying latency between hops, and modifying requests. Possible passive attacks include an operator collecting and storing data which passes through their Session Node(s) and logging all connections with other Session Nodes or clients in the network. 

\subsubsection{Network Adversary --- Passive Attacks}

Session’s threat model also considers a local network adversary such as an Internet Service Provider (ISP) or local network provider. This adversary can perform passive attacks such as monitoring traffic it relays, conducting deep packet inspection, or saving relayed packets for later inspection.

\subsection{Protections}

Session aims to provide the following protections against attackers within the scope of the threat model:

\textbf{Sender Anonymity:} A sender's Account ID (public key) cannot be linked to specific messages they send, except by the intended recipient. The IP address of the sender is unknown to all parties except the first node in the onion requests path. However, the first node does not know the intended destination or message contents.

\textbf{Recipient Anonymity:} The IP address of the recipient is unknown to all parties except the first node in the onion requests path. Again, the first node does not know the origin or contents of any messages.

\textbf{Data Integrity:} Messages are received intact and unmodified, and if messages are modified this can be detected.

\textbf{Storage:} Messages are stored and available until their specified expiry time.

\textbf{End-To-End Encryption:} All messages are encrypted only for the intended recipient(s) of a conversation.

\subsection{Out of Scope}

Attackers who are out of the scope of Session’s threat model may be able to break some of the protections Session aims to provide. 

\subsubsection{Network Adversary --- Active Attacks}

A network adversary could conduct active attacks including corrupting or rerouting packets, or adding delays for the purpose of denying service or correlating traffic. These attacks could compromise privacy and the accurate storage and retrieval of messages. These risks are primarily addressed by encrypting data and using onion requests to store and retrieve messages, making targeted attacks by network adversaries detectable.

\subsubsection{Global Passive Adversary}

A global passive adversary (GPA) that can monitor the first and last hops in an onion request path could use traffic analysis to reveal the IP address of a Session client and the destination node that client is talking to. This potential attack is a property of the onion request system; onion requests are a low-latency onion routing network, meaning that packets are forwarded to their destinations as fast as possible, with no delays or batching. This behaviour, while beneficial for user experience, makes traffic analysis trivial in the case of a GPA.

\subsubsection{Out of Band Key Discovery}

Users may compromise the pseudonymity provided by Session's public key-based account system. If a user associates their real world identity with their Account ID, other parties could discover metadata related to that Account ID.

\section{Foundations}

Session is built on the Session Node network, so it is important to understand what this network is, how it functions, and what properties Session derives from the network.

\subsection{Session Nodes}
\label{sec:session-nodes}

Many projects have attempted to establish decentralised permissionless networks. These projects have often found themselves struggling with a ‘tragedy of the commons’, wherein public servers required for the operation of the network are under-resourced and overused. This inadvertently causes the network to provide poor service to users, which discourages further use or expansion of the network. 

Conversely, those projects which succeed in creating large, public, permissionless networks face increased scrutiny into the parties which contribute to running the network infrastructure. This can be especially damaging when the operation of that infrastructure adversely affects the privacy, security, or user experience of network applications. For example, the Tor network faces constant questions regarding evidence of Sybil attacks from unknown parties attempting to run large sections of the public routing network, an attack which could be used to deanonymise users~\cites{Honey_Onions,Sybil_Attacks_Tor,Tor_Malicious_Exit}.  

Session seeks to avoid these concerns by using a different type of public access network: a staked routing and storage network called the Session Node network. This network layer operates on top of Arbitrum, an Ethereum Layer 2 blockchain \cite{Arbitrum}. Through the integration of a blockchain network, Session adds a financial requirement for anyone wishing to host a server on the network, and thus participate in Session’s message storage and routing architecture. 

Authorisation for a server to operate on the network is attained through the server operator interacting with the Session staking smart contracts. These contracts enable an operator to register a node by staking a defined amount of Session Tokens, which are bound to their Session Node public key.

This staking system provides a defence against Sybil attacks by limiting attackers based on the amount of financial resources they have available. The staking system also achieves two other goals which further reduce the likelihood of a Sybil attack.

Firstly, the need for attackers to buy or control Session Tokens to run Session Nodes creates a market feedback loop which increases the cost of acquiring sufficient tokens to run large portions of the network. That is, as the attacker buys or acquires more tokens and stakes them, removing them from the circulating supply, the supply of the Session Token is decreased while the demand from the attacker must be sustained. This causes the price of any remaining Session Tokens to increase, creating an increasing price feedback loop which correlates with the scale of the attack. 

Secondly, the staking system binds an attacker to their stake, meaning if they are found to be performing active attacks, the underlying value of their stake is likely to decline as users lose trust in the protocol, or could be slashed by the network, increasing the sunk cost for the attacker.

The other advantage of a staked blockchain network is that Session Nodes earn rewards for the work they do, paid as Session Tokens from the Session Node Staking Reward Pool. This system makes Session distinct from altruistic networks like Tor and I2P and instead provides an incentive linked directly with the performance of a Session Node. Honest node behaviour and the provision of a minimum standard of operation is ensured through a consensus-based testing suite (see section \ref{sec:testing-suite}). Misbehaving nodes face the threat of having their staked capital locked, while the previously-mentioned rewards function as the positive incentive for nodes to behave honestly and provide at least the minimum standard of service to the network.

\subsection{Onion Requests}
\label{sec:onion-requests}

The other foundational component of Session is an onion routing protocol, referred to as onion requests, which enables Session clients to obfuscate their IP address by creating a 3-hop randomised `path' through the Session Node network. Onion requests use a simple onion routing protocol which successively encrypts each request for each of the three hops, ensuring:

\begin{itemize}
  \item The first Session Node only knows the IP address of the client and the middle Session Node;
  \item The middle Session Node only knows the first and last Session Nodes but not the client or destination;
  \item The last Session Node only knows the middle Session Node and the destination IP address.
\end{itemize}

\begin{figure}[htbp]
\vspace*{+0.1in}

\begin{center}
\begin{tabular}{ccc}
\includegraphics[width=10cm]{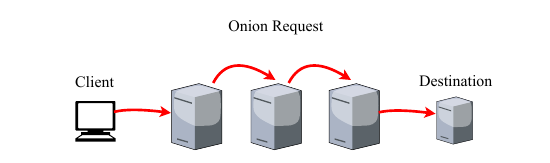}
\end{tabular}
\end{center}
\vspace*{-0.25in}
\caption{\textit{A client sends an onion request which hops via 3 Session Nodes to a destination, in this case, another Session Node}
\label{fig:OnionRequest}}
\end{figure}

\newpage

On startup, a Session client forms a `path' using three random Session Nodes $ (A, B, C) $ from their node list (see section \ref{sec:bootstrapping}). Note that the `path' is merely a logical abstraction, and although clients will reuse the same Session Nodes for subsequent onion requests, connections are not maintained between clients and Session Nodes in the path. The client's node list contains each Session Node’s IP address, storage server port, and Ed25519 public key. Clients use this information to create an onion request, encrypting each layer of the `onion' using the respective Ed25519 public keys of the three selected Session Nodes. This onion request is first sent to the storage server $A$; $A$ decrypts its layer of the onion, revealing the Ed25519 key of $B$ , the downstream node in the path, $B$. $A$ then initialises a ZMQ connection with $B$, if such a connection has not already been established. $B$ repeats this process, relaying the request to the final Session Node, $C$. $C$ decrypts the final layer of the onion, revealing the request's final destination; the request is then forwarded to the destination. The request's destination may be a Session Node's storage server, a Session File Server, or a Session Community server. The destination then encrypts a reply for the Session client and sends this reply back to the previous node, who sends it to the previous node, and so on until it reaches the Session client. 

Each onion request contains a payload; the payload describes what is being requested. For example, a Session client could be retrieving new messages, requesting an attachment, or sending a message. This payload is end-to-end encrypted for the destination. To encrypt the payload, the Session client generates an ephemeral X25519 key pair. This key pair, alongside the destination's converted Ed25519 key pair, is used to generate a shared symmetric encryption key. The payload is then encrypted with this shared symmetric key using either AES or the XChaCha20-Poly1305 algorithm, a variant of the widely-adopted ChaCha20-Poly1305 cipher suite \cite{Poly1305}. This payload can now be packed into the onion request, which contains necessary metadata, including encrypted routing information and the ephemeral X25519 public key.

\section{Building on Foundations}

The Session Node network provides an incentivised, self-regulating network of remote servers which provide bandwidth and storage space, and onion requests provide a straightforward anonymous networking layer. 

A number of services are built on top of this foundation to ensure Session provides the features commonly expected of modern messaging applications.

\subsection{Storage}

Message storage is considered an essential feature for any modern chat application. When a user sends a message, they expect the recipient to receive that message irrespective of whether they turn off their device after the message has been sent; users also expect successful message receipt when their device wakes from an offline state. Apps that run on decentralised networks typically cannot meet these expectations, because of the lack of incentive structures and, consequently, the ephemeral nature of clients and servers on such a network. Session is able to provide message storage through the incentivised Session Node network and its usage of swarms.

\subsection{Swarms}

Although the consensus of Session Nodes incentivises correct Session Node behaviour through rewards and punishments, these incentive models cannot prevent nodes going offline unexpectedly due to operator maintenance, software bugs, or data center outages. Therefore, for redundancy, a secondary logical data storage layer must be built on top of the Session Node network to ensure reliable message storage and retrieval. 

This secondary logical layer is provided by replicating messages across small groupings of Session Nodes called swarms. Any given Session Node's initial swarm is determined at the time of that Session Node’s registration, with the Session Node itself having minimal influence over which swarm it joins. Session Nodes deliberately lack agency over which swarm they join to prevent swarms being entirely made up of malicious or non-performant nodes. This contributes to the network’s self-regulating properties.

The composition of each swarm inevitably changes as the network evolves: some nodes leave the network and newly registered nodes take their place. If a swarm loses a large number of nodes it may "steal" a node from some other, larger swarm. In the unlikely event that the network has no swarms to steal from (i.e., every swarm is at \(N_{min}= 5\) nodes, the ‘starving’ swarm (a swarm with fewer than \(N_{min}\) nodes) will be dissolved and its nodes will be redistributed among the remaining swarms. Conversely, when a large number of nodes enter the network which would oversaturate existing swarms (i.e., every swarm is already at max capacity \(N_{max} = 10\)), a new swarm is created from a random selection of \(N_{target} = 7\) excess nodes.

The outcome of this algorithm is the creation and, when necessary, rebalancing of swarms of \(N_{\text{min}} \text{ to } N_{\text{max}}\)
Session Nodes which store and serve Session clients’ messages. 

The following set of simple rules ensure that Session Nodes within swarms remain synchronised as the composition of swarms changes:
\begin{itemize}
  \item When a node joins a new swarm, existing swarm members recognise the new member and push the swarm’s data records to the new member.
  \item When a node leaves a swarm, its existing records can be safely erased, with the exception of when the node is migrating from a dissolving swarm. In  this case, the migrating node determines the swarms responsible for its records and distributes them accordingly.
\end{itemize}

\subsection{Identity and Long-Term Keys}

Most popular messaging applications require the user to register with an email or mobile phone number in order to use the service. This requirement provides some advantages: sufficient standards (and associated cost) for account verification provide some spam protection; familiar identification mechanisms provide social network discoverability. However, such requirements represent a major privacy and security compromise for users. 

The use of a phone number as the basis for ownership of an identity key/long-term key pair weakens security against user accounts being compromised, such as in the cases of popular applications like Signal and WhatsApp. This weakness stems from the centralised management of phone numbers. Centralised service providers (i.e. telecommunications service providers) have the capacity to circumvent user control and assume direct control of specific users’ phone numbers. Widespread legislation already exists to compel service providers to exercise this power. Additionally, methods such as SIM swapping attacks, service provider hacking, and phone number recycling may be exploited by lower-level actors~\cites{Sim_Swap_Attack,WhatsApp_PIN,Signal_Twilio_Incident}.

Signal and WhatsApp put forward varying degrees of protection against these types of attacks. Signal and WhatsApp both send a "Safety numbers have changed" warning to a user's contacts if identity keys are changed. In practice, however, users rarely verify these details out-of-band~\cites{General_Authentication_Usability,Signal_Authentication_Usability}. 

Both Signal and WhatsApp also allow users to set a "registration PIN lock"~\cites{WhatsApp_2FA,Signal_Pin}. This protection means that an attacker (including a service provider or state-level actor) needs access to both the phone number and the registration PIN code to modify identity keys. However, this feature is off by default, difficult to find in the settings menu, and automatically disabled after periods of user inactivity. These factors all significantly reduce the efficacy of registration PIN locks as a protective measure against the security risks of phone number-linked accounts.

Using phone numbers as the basis for account registration also greatly weakens privacy for typical users. In most countries, users must provide personally identifying information such as a passport, driver's license or identity card to obtain a phone number---permanently mapping users’ identities to their phone numbers. These identity mappings are kept in private databases which can be queried by governments or the service providers that own them. There are also a number of web scrapers and indexers which automatically scrape phone numbers associated with individuals. These scrapers may target sources such as leaked databases, public social media profiles, and business phone numbers in order to link individuals to specific phone numbers. Since the primary method of initiating contact with a user in Signal, WhatsApp, or similar applications is to know the user’s phone number, this immediately strips away user anonymity---a significant concern for whistleblowers, activists, protestors, and other such users.

Other detractors of phone number account systems include: limiting the ability of a single user to establish multiple identities; and preventing high-risk users without access to a phone number from accessing the service.

Session does not use phone numbers or email addresses as the basis for its account system. User identity is established through the generation of an Ed25519 public-private key pair. This key pair is not required to be linked with any other identifier, and new key pairs can be generated on-device in seconds. This means that each key pair (and thus, each account) is pseudonymous, unless intentionally linked with an individual identity by the user through out-of-band activity.

\subsubsection{Account Restoration}

Because Session does not have a central server to keep records of users' accounts, the commonly expected user experience of account recovery using a username and password is not possible. Instead, users are prompted to write down their long-term private key upon account generation. This long-term private key is represented as a mnemonic seed phrase, referred to within Session as a Recovery Password. A user may use their Recovery Password to recover their account if their device is lost or destroyed. This enables the user's contacts to keep communicating with the same Account ID, instead of needing to establish contact with a new Account ID.

\subsection{Partitioning Identities}
\label{sec:swarm allocation}

We have described the grouping of Session Nodes into swarms to achieve redundancy for data storage. However, we also require a scheme to balance Session users across these groups, ensuring that swarms equally share the storage of offline messages. To do this, we divide the user-generated public key space into distinct, deterministic groupings and map each grouping directly to a swarm responsible for storing messages for users within that grouping.

Each swarm has a deterministically generated 64-bit number as its identifier. Once the identifier of every swarm on the network is known, we can create a mapping from clients' Account IDs to swarms. To do this, we first reduce a given Account ID to a 64-bit integer to match the number space of swarm identifiers. The key is then assigned to the swarm whose ID is numerically closest. This approach ensures that, given the list of current Session Nodes, a user can always determine their swarm without the use of a centralised resolver.

\subsection{Bootstrapping}
\label{sec:bootstrapping}

To route messages or use onion requests, each Session client needs an up-to-date list of all Session Nodes. On first launch, a Session client fetches this list over the clearnet from one of several hardcoded seed nodes. After first use, the Session client periodically queries multiple Session Nodes via onion requests (from its previously-acquired list) for an up-to-date list. Session clients can avoid being partitioned onto alternate (i.e. malicious) networks by updating their lists only when all queried Session Nodes are in consensus on a new Session Node list.

\subsection{Message Routing}

To route a message, the sender first needs to identify the Session Nodes in a recipient's swarm. This is done by obtaining the deterministic swarm mapping between the recipient's Account ID and the currently registered Session Nodes. Initially, this mapping is  requested from a random Session Node by the sender, but it is also updated as the nodes in a swarm change. In such cases, if the sender attempts to send a message to a node which is no longer in the recipient's swarm, they receive an error response from the Session Node. This response indicates that the Session Node is no longer part of the swarm and informs the client of the new nodes composing the swarm.

Given the swarm mapping, the sender creates a message and prepares the message to be sent to the Session Node network. The sender includes the necessary information for the message to be processed, including:

\begin{itemize}
  \item The Account ID of the recipient
  \item The message timestamp
  \item The message expiry time
  \item The namespace the message will be stored in
\end{itemize}

The sender then sends this message using an onion request to a random Session Node within the recipient's swarm. This Session Node will then propagate the message to the remaining nodes in the swarm, returning a signed success message to the sender if propagation is achieved. Each Session Node then stores the message until the specified expiry time.

\begin{figure}[htbp]
\vspace*{+0.1in}

\begin{center}
\begin{tabular}{ccc}
\includegraphics[width=16cm]{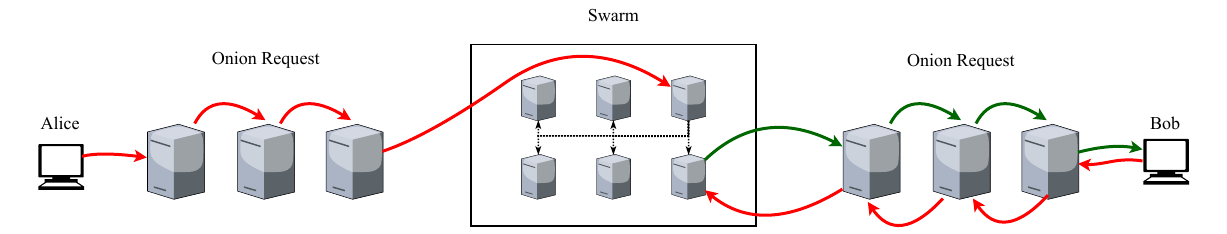}
\end{tabular}
\end{center}
\vspace*{-0.25in}
\caption{\textit{Alice uses an onion request to send a message to a Session Node within Bob’s swarm, the message is then propagated within the swarm. Bob later sends an onion request to retrieve said message.}
\label{fig:Asyncmessage}}
\end{figure}

\newpage

\subsection{Encryption}

So far, we have discussed both the transport and storage of messages. However, Session also requires message encryption in order to secure the contents of messages from being viewed by third parties. 

Given the requirements, the generally advocated position would be to use established message encryption protocols, like the Signal Protocol, which provide strong cryptographic guarantees against specific attack models \cite{Signal_Protocol}. However, these protocols are not suited for deployment in decentralised networks. This is partly because they were designed with a centralised server in mind, and thus make strong assumptions about establishing and synchronising shared states between clients. Other assumptions are also made, such as having permanent storage available for large pre-key bundles.

In the context of Session, 
strong synchronisation is difficult to maintain between clients, particularly when the only medium of synchronisation is via swarms---which do not provide message order or availability guarantees, do not provide long-term persistent storage (for example for large pre-key bundles), and remove messages once their expiry time has lapsed. This can cause messages to be missed when clients are offline for long periods of time.

Considering this, a new encryption protocol is required for decentralised messaging applications, such as Session.

\subsubsection{Session Protocol}

The Session Protocol is an encryption protocol specifically designed for deployment in decentralised networks, optimising to achieve high levels of security and reliability without requiring strong guarantees of client synchronicity and permanent storage mediums. 

In the context of one-on-one messaging, the Session Protocol provides a simple stateless end-to-end encryption protocol, and in the context of groups, it provides a protocol that can scale and tolerate faults caused by the constraints of the Session Node network. 

The protocol is designed to be easy to audit, implemented using existing, widely deployed, cryptographic libraries and suitable for the unique architecture of the Session Node network. The advantages and limitations of the Session Protocol are discussed further in
section \ref{sec:session-protocol-details}.

\subsubsection{Session Protocol Details}

This description covers the functions of the Session Protocol required for one-on-one chats. The function of the protocol for groups and communities is covered in sections \ref{sec:groups} and \ref{sec:communities} respectively. 

To construct and send a valid message in a one-on-one chat using the Session Protocol, the sender first creates a message $M$ containing a number of data fields, including the plaintext message, and the senders Ed25519 public key. $M$ is then serialised and padded to the next 160 byte multiple to prevent extraction of the exact message size. $M$ can now be signed given the sender's Ed25519 private key using the Ed25519 signature scheme. Once signed, the sender's Ed25519 public key and the signature are then appended to $M$.

After signing the message $M$, the sender generates an ephemeral X25519 key pair. This key pair, alongside the recipient's X25519 key pair, is used to generate a shared symmetric encryption key. The message $M$ is then encrypted with this shared symmetric key using the XSalsa20-Poly1305 algorithm. This encrypted blob can now be packed into a message, which contains necessary metadata specifying the recipients X25519 public key and the ephemeral X25519 public key. Though both Ed25519 and X25519 keys are used, only one private key is required; the X25519 keys are derived from the recipient's Ed25519 keys.

The usage of both Ed25519 and X25519 keys in this process may create some confusion. The public X25519 key is derived from the sender's Ed25519 key pair; this public X25519 key represents the user's Account ID. Historically this was done to maintain compatibility with the Signal Protocol which also uses X25519 keys in this way.

Once a message has been generated and packed it will be sent to the recipient's swarm via an onion request specifying  the expiry time and timestamp of the message. The recipient will authenticate retrieval of the message and reverse the process, decrypting, determining, and authenticating the sender of the message as it does so.

\section{Groups \& Communities}

Increasingly, instant messaging applications are places for groups and communities to gather, rather than simply being used for one-on-one conversations. This has led to widespread use of group chats, channels, and similar functionality in messaging applications~\cites{Group_Chat_Stats,Group_Chat_Usage}. The most popular messaging applications support group chats, but the levels of encryption and privacy provided to users in these group chats is often unclear or non-existent. By default, group chats in applications such as Telegram and Facebook Messenger only support transport encryption, rather than end-to-end encryption. Even those applications which do support end-to-end encryption in group chats (e.g. Signal and WhatsApp) still use central servers to store and disseminate messages.

The deployment of encrypted group chats in Session focuses on meeting scaling and encryption requirements.

\pagebreak

\subsection{Scaling}

There are two main approaches to sending messages in a group chat: server-side fanout and client-side fanout. This choice has a significant impact on the scalability of the group chat.

In client-side fanout, the client pushes their messages to each recipient device or swarm. Client-side fanout is preferable in some cases, as it can be achieved in peer-to-peer networks and does not require the establishment of a central server. However, client-side fanout can prove burdensome on client bandwidth and CPU usage as the number of group members increase---a factor which proves particularly problematic for mobile devices. Additionally, client-side fanout cannot deliver messages to offline devices, a major issue if a recipient device is not online or connected to the network when a message is sent. 

\begin{figure}[htbp]
\vspace*{+0.1in}
\begin{center}
\begin{tabular}{ccc}
\includegraphics[width=6cm]{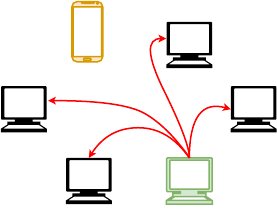}
\end{tabular}
\end{center}
\vspace*{-0.25in}
\caption{\textit{Client sends message using client-side fanout: The sender (green 
computer) sends messages to other clients (solid red lines) but is unable to send to an offline device (yellow phone)}
\label{fig:clientside}}
\end{figure}

In server-side fanout, the client first pushes their message to a server, the server then disseminates the message to each recipient device or swarm. The recipient clients may also fetch the message from the server at a later point in time, which is more efficient for larger groups.

\begin{figure}[htbp]
\vspace*{+0.1in}
\begin{center}
\begin{tabular}{ccc}
\includegraphics[width=6cm]{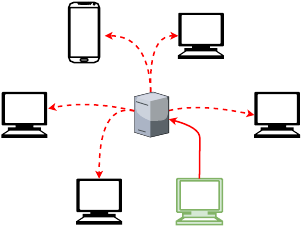}
\end{tabular}
\end{center}
\vspace*{-0.25in}
\caption{\textit{Client sends message using server-side fanout: The sender client (green computer) pushes the message to the server (solid red line) and the server distributes the messages to recipient clients (dotted red lines)
}
\label{fig:serversidefanout}}
\end{figure}

\newpage
\subsection{End-To-End Encryption}

Another factor impacting group chat scalability is the specific end-to-end encryption implementation.

The most naive solution to build encrypted group chats in Session would be to leverage the existing pairwise sessions we can create for one-on-one conversations. To send a message to a group chat, a pairwise session would be started with every member of the group, and each message would be individually encrypted for each participant. This provides the group chat with the same guarantees possessed by standard one-on-one communications using the Session Protocol. However, this would come at the cost of requiring the payload to be encrypted, sent, and stored \(N-1\) times, where \(N\) is the number of members in the group. This process is burdensome for low-powered devices when participating in large group chats. This would also incur a non-trivial increase in storage load on the Session Node network.

One way to improve scalability in group chats is to adopt a shared encryption key for all group members. This key would be shared in a pairwise message to members joining the group. When a member leaves or is expelled, the administrator creates a new shared encryption key, re-encrypts it, and sends it to the remaining \(N-1\) members, effectively rotating the key to prevent former members from accessing new messages. In this system, messages only need to be encrypted once for the group in most scenarios.

This scheme is effective in small to medium-sized group chats where the membership set changes infrequently. However, it can be impractical in larger groups where users frequently leave (or are otherwise removed from) the chat.

\subsection{Other Considerations}

\subsubsection{Group Size}

It may be possible to create large encrypted groups which scale well even when members are added and removed frequently \cite{MLS_Protocol}. However, as groups become larger, it becomes increasingly likely members will leak or otherwise share conversation contents directly, undermining all cryptographic protections. Thus any design should weigh the overhead of end-to-end encryption schemes against the size of the group. 

\subsubsection{Metadata Protection}

Information about a group chat, including the public keys and IP addresses of members and administrators, should be kept private from all third parties, as public availability of information about the relationships between public keys and IP addresses significantly reduces privacy. This metadata protection should also extend to the group name, avatar and description.  

\subsection{Group Types}

With the above considerations in mind, the Session Protocol defines two different designs for the encryption and scaling of group chats. Which design is appropriate depends on the intended size and use of the group and is chosen when the group is created. `Groups' support communication among 3 to 100 participants. Participant sizes beyond 100 require the use of `communities'. However, it should be noted that `communities' can also accommodate participant sizes fewer than 100.

\subsubsection{Groups 3 - 100 Members}
\label{sec:groups}

To initialise a group, the creator of the group selects a number of members from their existing conversations or directly via their Account IDs. The creator then generates a group identity key pair, a group encryption symmetric key, and a pair of authentication values for each invited member.

Each of these generated elements has a specific purpose within the group.

\begin{itemize}
  \item The group identity key pair is used to identify the shared swarm where members send and retrieve messages and to authenticate the group administrators' actions.
  \item The group encryption symmetric key is employed to encrypt and decrypt messages within the group, ensuring secure communication.
  \item Authentication values are provided for non-administrator members to authenticate message sending and retrieval from the group's swarm (see section \ref{sec:Authentication}).
\end{itemize}

After generating these elements, the creator encrypts the group encryption symmetric key with the Account ID of each new member, creating an array of encrypted payloads. Each payload contains the group encryption symmetric key, individually encrypted for a member. This array is then stored in the group's dedicated swarm and periodically refreshed to ensure availability.

The creator then sends group invite messages to each invited member. This invite includes the group's name, the group identity public key, a digital signature using the group identity private key, and the authentication values for each member.

Given the group identity public key, members can now determine the specific swarm that stores messages for their group (as detailed in section \ref{sec:swarm allocation}). Using their authentication values, members authenticate themselves to a Session Node within the swarm and retrieve the array of encrypted payloads. Members then trial decrypt each payload until they successfully decrypt the group encryption symmetric key. 

Onion requests are used for transmitting messages to and from the shared swarm.

\textbf{Group Administration}

The creator of a group acts as the first administrator of the group. Group administrators may install additional administrators by sharing the group identity private key. Administrators can remove group members, delete members' messages from the group, and delete the group. Members have the ability to read and write group messages, delete their own messages, and leave the group.

\textbf{Group Encryption}

Users encrypt and decrypt messages for the group using the established group encryption symmetric key, messages are encrypted using the XChaCha20-Poly1305 algorithm.

\textbf{Group Key Rotation}

To maintain secure communication within a group, it is important to prevent a member who has left the group, either voluntarily or due to being kicked by an administrator, from decrypting any future messages. This necessitates the rotation of the group encryption symmetric key. If a member is kicked, the administrator who took the action must generate a new symmetric encryption key for the group and distribute it to the remaining members. This is achieved in the same way as the group encryption symmetric key is distributed to the initial group members.

This process is complicated if multiple administrators are attempting to remove members at the same time, or a member leaves and an administrator or multiple administrators have linked devices. In these cases, individual administrator devices will race to update the group encryption key, potentially causing conflicts. These conflicts cannot be resolved by the Session Node network, as Session Nodes do not have the required visibility to merge conflicting updates. Instead, conflicts are resolved by individual administrator devices. Each client assigns a sequence number for their update. If clients send different updates with the same sequence numbers, then individual clients will merge these updates together, increment the sequence number, and resend them to the swarm. Storage servers hash all messages, deleting duplicates to prevent update loops.

\subsubsection{Communities 100+ Members}
\label{sec:communities}

Large groups run into significant scaling issues when members leave the group, as encryption keys must be updated and refetched by the entire group---an inefficient process when there may be hundreds or thousands of members. Additionally, as previously addressed, the usefulness of end-to-end encryption in very large groups is unclear, as a single malicious group member or compromised device is able to compromise the privacy of an entire group. In large groups, this is extremely difficult to protect against regardless of the encryption deployed. Thus, it makes sense for large groups to conform to a different standard than regular groups. In Session, these large groups are called communities. Communities provide transport-only encryption, protecting users against network adversaries but providing comparably weak protection against server-side attacks.

To balance the risk of such attacks, Session Communities do not store messages on the Session Node network. Instead, communities require self-hosting using open-source software~\cite{Session_Community_Server}. All messages and attachments stored in communities are fetched and posted using onion requests forwarding to the IP address or domain name of the Session Community Server. This preserves network-layer anonymity for all community participants.

Session Community Servers may have several rooms. Each room displays as a separate conversation in the Session client. This allows server operators to operate many rooms for many different topics of conversation without requiring the deployment of additional servers.

\textbf{Community Administration}

Administration of communities is comparably more complex than that of groups. The community's server operator is the original administrator, and they are able to add new administrators and moderators. Community administrators have the ability to modify members' read, write, upload, and access permissions on a per room basis. Administrators also have the ability to delete other users' messages and attachments from the room to keep conversations orderly. 

\textbf{Session Community Server Encryption \& Authentication}

Encryption of requests to a Session Community Server is handled in the same way that Session one-on-one communications are handled in the Session Protocol, with the main difference being that the targeted Account ID is the Session Community Server public key, which is communicated out of band in the Session Community URL. 

\begin{figure}[htbp]
    \centering
    {\scriptsize \urlstyle{same} % Temporarily set URL font to scriptsize
    \url{https://open.getsession.org/session?public_key=a03c383cf63c3c4efe67acc52112a6dd734b3a946b9545f488aaa93da7991238}}
    \caption{\textit{Example of a Session Community Server URL}}
    \label{fig:opengroup}
\end{figure}

Authentication to the Session Community Server is similar to authentication in one-on-one conversations, with one main difference. Instead of each user using their long-term Session private key pair to sign messages, they use a deterministic derived key. This ensures users do not leak their real Account ID to large communities, as this could be a vector for spammers to collect users' Account IDs. 

Users derive their blinded Account ID by first hashing the Session Community Server public key alongside their Account ID,
$K=Hash(Account ID || Server\_PublicKey)$
then reducing this hash $K$ to an Ed25519 scalar $k$. With this scalar, we can compute a derived Ed25519 key pair $A' = kA$  and $a' = ka$ where $A$ and $a$ are the user's Public Account ID and Private Account ID respectively. 

Once this key pair is generated, users sign messages using $a'$, with $A'$ as their publicly visible blinded ID. Signatures are validated as normal by checking the signature appended to the message was signed under $A'$.

Beyond just hiding the Account ID of a user, this scheme offers a way for users to initiate encrypted contact with each other using their blinded IDs and  recognise links between a user's blinded ID and Account ID if they have had previous contact in a closed group or one-on-one conversation. 

The former property is achieved through the mediation of message requests by the Session Community Server (see section \ref{sec:message requests}). Typically message requests are encrypted for the recipient's Account ID and sent to the corresponding swarm, however this is not possible in the case of ID blinding. In this case, a user may instead encrypt a message request for a recipient's blinded public key. This message, containing the sender's Account ID and a message, is sent to the Session Community Server and received as a message request the next time the recipient polls for messages from the server. The recipient may accept the message request by sending a message encrypted for the sender's Account ID and to the sender's swarm. Once this step is complete, the Session Community Server's involvement is no longer required to relay messages between the users. Importantly, relaying the initial blinded message request through the Session Community Server allows the server to apply rate limiting and other controls on specific blinded ID's to prevent spam.

The latter property is achieved as users of a Session Community who have previously interacted may calculate $K$ for one another using the Session Community Server public key and each other's real Account IDs $A$, subsequently they can calculate $k$, and since $A' = kA$, they can recognise each other's blinded public keys. As this mapping is possible, users can initiate chat requests with existing contacts directly without passing messages through the Session Community Server.

\section{Additional Layers}

Combining the aforementioned elements, we have constructed a lightweight, secure messenger. This messenger allows users to communicate both synchronously and asynchronously in one-on-one, group and community contexts without using a central server or exposing metadata such as IP addresses or phone numbers. However, users expect additional features beyond these foundational elements from modern messengers. The implementation of these additional features requires further extension of the Session design, which will be explained herein.

\subsection{Multi-Device}

Modern messaging applications are expected to sync data (message histories, contact lists, etc.) across multiple devices, ensuring  users are able to, for example, swap between a laptop and a mobile phone and continue their conversations. 

Devices received messages can be easily synced in Session by sharing the long-term private key associated with an Account ID between two devices, this will allow both devices to fetch one-on-one messages from the same swarm, and send new messages signed under the same Account ID.

Sent messages can be synced by ensuring a duplicate of every sent message is deposited to the sender's own swarm. This duplicate can be retrieved and decrypted by any linked device. 

The ease of linking devices in Session is mostly a consequence of the statelessness of the Session Protocol in a one-on-one context. However, some state must be preserved in group chats and to retain profile information.

\newpage

\begin{figure}[htbp]
\vspace*{+0.1in}
\begin{center}
\begin{tabular}{ccc}
\includegraphics[width=16.5cm]{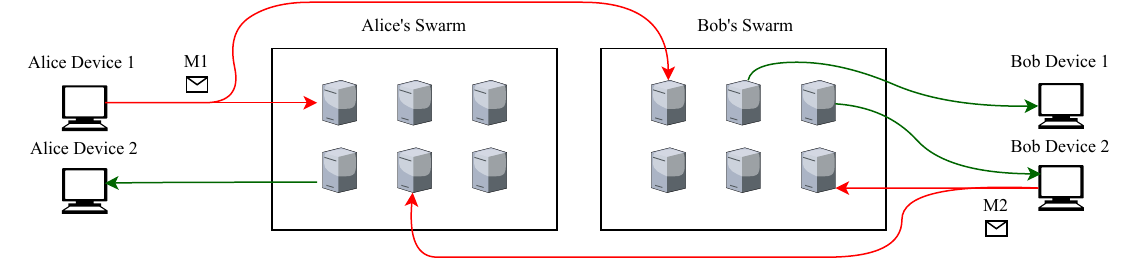}
\end{tabular}
\end{center}
\vspace*{-0.25in}
\caption{\textit{Alice sends Bob a message M1 via device 1, M1 is deposited in Bob's swarm and Alice's swarm, Alice's second device receives M1 as a sent message, Bobs two devices receive M1 via Bob's swarm, Bob responds via device 2 with message M2, which is deposited in Alice's and Bob's swarms}
\label{fig:friendrequestmultidevice}}
\end{figure}

\subsubsection{Configuration messages} 

Configuration messages contain content which can aid in linking or restoring accounts. Configuration messages generally include profile information, such as user nickname, profile image, contacts lists, lists of communities joined, and current group encryption keys. Configuration messages are encrypted against the user's Account ID and are stored in the swarm belonging to the user.

To ensure configuration messages are kept up to date and do not expire, clients update old configuration messages with refreshed information each time a change occurs (or at least once every 7 days), at the same time extending the expiry time of the configuration message. This ensures configuration messages are available if at least one of the user's client devices has been online recently. 

\subsubsection{Namespaces}

Clients are able to send several different types of Session messages, including visible messages and configuration messages. Some types of messages, like configuration messages, need to be fetched much less frequently than visible messages. To account for this, Session sorts messages into a limited number of types using logically separated inboxes called "namespaces". Each Session client tags outgoing messages with the namespace where the message should be stored. Session clients are then able to fetch messages from specific namespaces. 

This does create additional metadata which is visible to Session Nodes in the user's swarm. However, namespace types are limited to configuration message types and visible messages. These types are naturally more likely to have temporal or size signatures due to consistent client behaviour. Thus, explicitly tagging them is unlikely to have a significant effect on user privacy.

\subsubsection{Syncing Communities}

As derivation of blinded Account IDs is a deterministic process based on the user's long-term key pair and the Session Community Server public key, any linked device will generate the same blinded Account ID for each Session Community Server. This allows linked clients to identify messages previously sent by other devices in communities and sign new messages originating from their shared blinded ID. 

\subsection{Attachments}

Although Session Nodes have the ability to store data on behalf of clients, this responsibility only extends so far. Requiring Session Nodes to store attachments---which can easily be orders of magnitude larger than messages---would place an undue burden on the Session Node network.

With this in mind, a logical solution is for Session to interface with an untrusted, centralised server which stores data obliviously. As long as the central server cannot know the contents of files, or who is storing and requesting the files, this system does not cause any metadata leakage.

This is achieved by first padding each attachment to fit within a fixed number of constant sizes between 0 and 10 megabytes, then encrypting the attachment with a random symmetric AES key. The sender uploads the encrypted file using an onion request, and in response, the Session File Server returns a link to the piece of content via the onion request path.

Once the sender obtains this link, they send a message to the recipient via an existing pairwise session. This message contains a link to the content and the decryption key. The recipient uses an onion request to pull the encrypted attachment from the centralised file server and decrypts it locally using the decryption key provided by the sender.

By default, all Session clients use a Session File Server run by the OPTF for attachment sending and storage \cite{OPTF}. Since attachments are not considered a core feature of Session, this design is in keeping with Session's design principles. The Session File Server is fully open-source, with setup instructions provided such that users are able to set up their own file server~\cite{Session_File_Server}. In the future users should be able to specify in the Session client which Session File Server their client should use for attachment sending functionality. This is important both for providing users with choice and control, and ensuring the continued usefulness and functionality of Session if the OPTF was unable to maintain the default Session File Server.

\subsection{Client-Side Protections}

Secure messaging applications have typically focused their development efforts towards providing protections against network and server level adversaries, which has led to new advances in encryption and metadata protections. However, when interviewing high risk individuals, researchers found client-side privacy and security protections are some of the most-requested features. High-risk individuals may not be focused on protecting themselves against global adversaries, but instead against a small nation state, or corporate entity~\cite{High_Risk_User_Study}. For these individuals, endpoint compromise, device seizure, and forced disclosures are described as the biggest risks. To better mitigate these risks, Session implements a number of client-side protections which allow users to better manage the security of the Session app on their device.

\subsubsection{Deletion}

Granular message and data deletion controls are important for users who are likely to have their devices physically seized. Session implements standard features like disappearing messages, with options to delete messages a pre-specified time after they are sent, or a custom time after being viewed. Session also offers the ability to wipe all data from your swarm and device in a single action. Future releases will include duress codes and remote deletion options to further strengthen client-side protections.

\subsubsection{Message Requests}
\label{sec:message requests}
The first message or group invitation a user sends to an unknown contact will always be interpreted by the recipient as a message request. Message requests shield users from potential spam by sorting messages into a special folder separate from the regular conversations screen. Users can then decide whether they want to accept or deny further messages from the user. If accepted the conversation or group moves from the message requests section of the application into the main conversations screen.

\subsubsection{Pseudonyms}

High-risk users (such as whistleblowers) often need to create accounts without using personal identifying information (e.g. phone numbers and email addresses). Session account creation only requires generation of a public-private key pair, making it trivial for users to establish multiple pseudonyms without needing to link their account to personal information which could be used to deanonymise them. 

\subsection{Human Readable Usernames}

\subsubsection{Session Name Service}

Account IDs are 66-character hexadecimal addresses; they are a representation of the user's long-term public key. To add another user as a contact, you must know their Account ID. Typically users add a new contact by scanning a QR code representing this public key, or by directly receiving the full 66-character key out of band (typically via relatively insecure communications channels such as an unsecured messenger, SMS, or email).

This is not ideal from a user experience or privacy standpoint. Preferably, users would be able to reserve a short, human-readable username and share this username with friends, while still maintaining their security and privacy. Session Name Service (SNS) allows users to register a mapping between their Account ID and a shorter, human-readable string of text. This mapping is stored on the Arbitrum blockchain through interaction with a name registration contract; this contract burns a small amount of Session Tokens to register a name. This transaction requirement exists to guard against frivolous name registrations.

By storing these mappings in the Arbitrum blockchain, which is accessible to every registered Session Node, we can ensure each Session Node has a copy of every Account ID-SNS mapping, and that those mappings cannot be modified by unauthorised parties onchain.

When a Session client wants to add a new contact using SNS, the client connects to a number of random Session Nodes via onion requests and asks for the Account ID associated with that SNS. Name resolution will only succeed in the case where all Session Nodes return the same response. The safety of name resolution then depends on whether an attacker can register malicious nodes which return incorrect mappings. This attack is limited by the inherent Sybil attack resistance of the Session Node network. 

\subsection{Testing}
\label{sec:testing-suite}

Session Nodes are rewarded for providing services to the network in an honest and consistent manner. Consequently, dishonest Session Nodes must be prevented from refusing to store messages for the network while continuing to collect rewards. This is accomplished through Session Node testing, a network-level system of peer policing.

\subsubsection{Session Node Reachability Testing}

To ensure Session Nodes are providing service to the network, Session Nodes periodically test each other and report results in a decentralised policing system. This begins with each Session Node randomly sorting their own Session Node list. Session Nodes then work their way through the list testing other nodes by attempting to establish connections to that nodes storage server and Lokinet instances. If a connection is established, a node is marked locally as passing the that test. Session Nodes also publish uptime proofs through the p2p layer periodically. Session Nodes listen to each other's uptime proofs, and if a node does not send an uptime proof for longer than 2 hours the node is presumed offline. This is also taken into account during testing. 
 
Every Session Node periodically participates in a process to generate pseudo-random unpredictable data using a RANDAO style commit reveal scheme \cite{RANDAO}. This data is shared amongst nodes. Using this data, Session Nodes sort themselves into testing quorums of 10 nodes. Testing quorums collectively report their local testing results on a deterministically selected set of 50 Session Nodes. If any Session Node is reported as non-functional by more than 7/10 of the testing quorum, then a jointly signed transaction is emitted by one of the nodes, which will lead to the offending node being decommissioned or deregistered from the Session Node network.

\section{Session Protocol \& OTR Properties}
\label{sec:session-protocol-details}

Off-the-record (OTR) enabled messaging protocols generally seek to provide a number of cryptographic guarantees to conversations, namely Perfect Forward Secrecy, cryptographic deniability, and occasionally, self-healing. The most popular and well deployed OTR messaging protocol is the Signal Protocol, which is deployed to over a billion devices across Signal, WhatsApp, and Facebook Messenger. It is important to discuss how the Session Protocol compares to OTR protocols like the Signal Protocol so that threat models and practical cases can be understood. 

\subsection{Perfect Forward Secrecy}

Perfect Forward Secrecy (PFS) is a property of OTR messaging protocols which protects old conversations or messages from being compromised when long-term key information is exposed. This is achieved via a mechanism called key ratcheting, which periodically derives new shared ephemeral encryption keys using Diffie-Hellman key exchanges. These keys are deleted from the device each time new keys are derived, meaning messages encrypted using old keys will no longer be able to be decrypted, regardless of whether the device's long-term private key is compromised.

In the case of one-on-one conversations, the Session Protocol is stateless and requires no ratcheting of key material. It encrypts for the long-term public key (Account ID). These properties of the Session Protocol mean that in one-on-one conversations PFS is not provided.

\subsubsection{Practical Application}

Assuming a chat application properly manages key pairs, the only way a long-term key compromise should occur is via local device compromise. This detail limits the cases in which PFS is applicable---if an attacker has device access, plaintext messages in most cases can be pulled directly from the application database. 

However, PFS is useful at securing message contents in some circumstances.

Specifically, if the attacker has previously intercepted messages at the network level, stored them, and has local device access, PFS would prevent the attacker from reading those messages, as the encryption keys  for the intercepted messages would have been discarded by the application database. Without PFS, the attacker would be able to decrypt the network scraped messages with the user's long-term key. 

It is important to note sophisticated attackers with full network and device access are likely to perform easier and more damaging attacks, such as installing monitoring software on the device, reading future messages, and compromising current contacts---which PFS does not protect against. Further complicating this specific attack, the Session Node network deploys protections to prevent collection of messages at the network level, which is a prerequisite to perform this attack (see section \ref{sec:scraping}). 

\subsection{Cryptographic Deniability}

Cryptographic Deniability, or deniable authentication, allows for both parties in a conversation to authenticate the origin of messages they receive during a conversation, but prevents either party from proving the origin of the message to a third party post-conversation. This property is provided in the Signal Protocol through the derivation of per-conversation shared encryption keys which are used for encryption and authentication. Messages are not signed with a long-term identifiable key.

Since in all uses of the Session Protocol, messages are signed with the long-term private key associated with an Account ID, cryptographic deniability is not provided.

In practice, cryptographic deniability is often disregarded when it comes to court cases or media reporting. For example, cryptographic deniability was used unsuccessfully as a defense in a court case involving the communications of Chelsea Manning and Adrian Lamo \cites{High_Risk_User_Study,Real_World_Deniability}. Instead, courts often rely on screenshots of conversations from either a chat participant or seized devices to establish real world identities.

Additionally, deniability fails to provide protection if a chat partner or device is compromised during the conversation. In such a case, the compromised participant or device can prove which device or chat partner sent which messages post-conversation.

Although Session signs messages with the long-term private key associated with the user's Account ID, signatures are immediately deleted after validation. This means an unmodified client will never have access to signatures which prove messages were signed by a specific party. Thus, an attack on deniability would require modified client software to store signatures before deletion or scrape messages off the network before the signature was deleted.

\subsection{Groups}

The Session Protocol handles group encryption differently from one-on-one conversations. In groups, whenever a user is invited by an administrator, the administrator can choose whether to rotate the group encryption symmetric key. Additionally, whenever a user leaves or is kicked from the group, the group encryption symmetric key is automatically rotated. Once these keys are no longer needed, they are purged from the device. This means group conversations inherit a level of natural forward secrecy as membership evolves.

\section{Scraping}
\label{sec:scraping}

As Session uses onion requests to route requests to destination swarms, useful network scraping attacks are not possible at the Session Node level during message transit due to messages being encrypted with additional layers of ephemeral onion-based encryption. 

Instead, the only parties that receive the encrypted request intended for the recipient are the Session Nodes comprising the recipient's swarm. However, if the retrieval of encrypted messages from a swarm was unauthenticated, trivial network scraping could occur. This would allow any non-Session Node party to request all messages tagged to a specific Account ID which they do not own. To protect against this, individual Session Nodes in each swarm require authentication from users who request messages. 

\subsection{One-on-one Conversation Authentication}

Users' clients will regularly poll Session Nodes in their own swarm for new messages. Each of these requests requires a signature from the user's Account ID private key. The data being signed is the type of request, the namespace of the request, and a timestamp. The Session Node will validate the type of request, check that the namespace requires authentication and check whether the timestamp is within an acceptable range to limit replay attacks. If all checks pass, the Session Node will validate that the Account ID used to produce the signature matches that of the message being fetched. Once this process is complete, the Session Node will send a response back to the client. 

\subsection{Community Authentication}

Communities use blinded Account IDs to authenticate Session users. A blinded Account ID is the public key of a key pair that is derived from the user's Account ID private key and the Session Community Server's public key.

When a user joins a community, they produce signatures using their blinded key pair. Before messages are disseminated to the community, the Session Community Servers validate whether the signatures match the claimed blinded ID. Clients also validate these signatures locally to ensure authenticity in the case of a dishonest Session Community Server.

\subsection{Group Authentication}
\label{sec:Authentication}

Preventing scraping in a group context is more challenging since the swarm of Session Nodes storing messages for a group only have visibility of the long-term group identity public key. This means Session Nodes can only validate requests for the group when they are signed under the group identity private key, which is only given to group administrators. Group administrators sign every request to the Session Node network under this key and Session Nodes validate these signatures, just as would occur in a one-on-one conversation. As a consequence of the group identity public key being static, once a member is made an administrator, they can scrape and decrypt future messages even if they leave or are kicked from the group. This detail is made clear at the UI level when adding new administrators to the group. 

Unprivileged members of a group require a different method of authentication. When a new member of the group is added by an administrator, the administrator will generate a pair of values: a token and a signature. The token contains a blinding factor, permission flags, and a network prefix. Similarly to blinded keys used in Session communities, a blinded public key $A'$ is generated by the administrator such that $A' = kT$, where $k$ is the blinding factor and $T$ is equal to the absolute value of the user's public Account ID. Permission flags $f$ represent capabilities such as the capability to read, write or delete messages in the group. The final piece of information in the token is the network prefix of the account (03 for groups). The administrator will sign over the token with the group identity private key and will then send the signature and token to the new member of the group. 

Upon receiving these credentials, the new member calculates their corresponding blinded private key, represented by $a' = kt$, where $t$ denotes the private scalar linked to the user's public Account ID.

To fetch group messages, the member sends a request to a Session Node within the group's swarm. This request will contain a signature from the blinded private key $kt$, the previously provided token (excluding $T$), and the administrator's signature over that token from the group identity private key. 

The Session Node verifies the administrator’s signature of the provided token using the group identity public key. This proves to the Session Node that the blinded token has the permissions within the group which are enumerated in  the flags $f$. Subsequent to this verification, the Session Node confirms the member's signature under the blinded public key $A'$, proving that the member possesses the blinded private key $a'$.   

When a member is kicked or leaves, a group administrator will add the removed members' blinded public key $A'$ to a revocation list which consists of the keys of the last 50 members who have been kicked or left the group. This key revocation list is stored unencrypted in the swarm such that the Session Nodes can read revoked tokens. When Session Nodes validate requests, they refuse requests signed with a revoked token.

This mechanism accomplishes a few key goals 
\begin{itemize}
  \item Protecting the social graph of the group: Session Nodes never become aware of the actual Account IDs involved in the group.
  \item Non-interactivity between administrators: members can be added and removed from groups without interaction between administrators who may be offline.
  \item Scraping protection: non-administrators can be blocked from retrieving messages.
\end{itemize}

\section{Attacks \& Future Work}

\subsection{Spam} 

Since Session `accounts' are simply public-private key pairs, they can be generated en masse with relative ease. Client-based Sybil attacks are possible to execute, making moderation of communities and even personal conversations challenging.

Protections must be designed to be minimally intrusive to real human users---including those utilising relatively low-powered mobile devices---while preventing automated attacks from malicious actors possessing large amounts of compute, storage and bandwidth. Several approaches may be possible, but each have distinct cost benefit profiles. These approaches include: 

\begin{itemize}
  \item Proof of work applied on a per message basis.
  \item CAPTCHA schemes.
  \item Limiting connections at the first hop of an onion request based on IP address.
  \item Community based moderation tools and bots to detect and facilitate the removal of spam.
\end{itemize}

Investigation and implementation is ongoing. 

\subsection{Enhanced Session Node Testing}

Currently the network only tests the reachability of individual Session Nodes. The network does not test whether individual Session Nodes are storing user messages. Presently, this has a relatively low impact, as storage requirements for Session are trivial. However, as the network grows and storage on Session Nodes increases, Session Nodes may seek ways to reduce storage costs by only responding to reachability tests and refusing to store user data. Thus, designing deeper testing of Session Node storage behaviours is of great import.

If Session Node \(A\) and \(B\) both belong to the same swarm, they are expected to store the same database records. One approach would be for Node \(A\) to select several records from its database (verifying that the messages indeed belong to the current swarm) and send a test request to node \(B\), containing only the record hashes, the current block height \(h\) and a randomly generated nonce \(n\). \(B\) is expected to respond with a new hash proving their knowledge of the record data.

When \(B\) receives a test request, it checks the specified height \(h\), and confirms that \(h\) is within some reasonable boundary. It then tries to retrieve the requested records from its own database. When the requested records are obtained, \(B\) hashes the data of those records locally along with the nonce \(n\) and sends this response to \(A\). \(A\) can validate this proof by performing the same actions locally. Importantly, this proof cannot be generated without the nonce or record data or by another Session Node. If \(A\) does not receive the response within some acceptable time window (or it is incorrect), \(A\) reports \(B\) in a testing quorum as having failed the test. In cases of repeated failures, the Session Node network may choose to decommission or deregister \(B\).

\subsection{Lokinet Integration}

Session currently uses the onion requests protocol to send and receive data from the Session Node network. This protocol has a number of limitations, one such limitation being  the protocol is designed as a stateless request response protocol, meaning specific requests must be constructed for every piece of information the Session client requires or provides. Therefore, the Session client must retrieve all information using polling, which is inefficient and slow; in comparison with a stream based protocol, where connections can be held open for both server and client to push events. Another limitation of this stateless protocol is that it places requirements to limit the maximum amount of data that can flow in a single request between Session Nodes in order to ensure memory is not clogged while Session Nodes handle a request. Currently, this limit is set at 10MB which consequently limits the maximum size of attachments. 

Lokinet integration is designed to resolve both of these limitations. Lokinet is a network layer onion routing protocol, which can facilitate stream based protocols such as QUIC, allowing clients to connect to Session Nodes and listen for events related to their Account ID or push new messages without needing to repeatedly poll \cite{Lokinet}. Since streaming support is provided, large requests containing files can be broken up into packets and streamed to the destination without any size limit.  

Lokinet support for stream-based connections extends beyond just client to Session Node communication, Session could theoretically allow clients to connect directly to each other to stream data without requiring intermediary storage. This would be useful for maintaining anonymity for voice and video calls and for the transfer of large files between online clients.

\subsection{Wallets}

Session already provides self-sovereign ownership of a cryptographic key pair. Although this key pair is usually used to sign, encrypt and decrypt messages, it is also possible to derive an EVM-compatible wallet from this key pair. With appropriate integration, this would allow the user to use a lightweight EVM wallet integrated directly into the Session app. This wallet would make it easy for users to quickly transfer value inside Session while conforming to a commonly understood messaging user experience. 

\subsection{Social Network Discovery}

As previously discussed, common messaging applications like WhatsApp, Facebook Messenger, and Signal typically require the user to provide their phone number or email address upon registration. These apps are then able to access the user's pre-existing contact lists and social networks to discover which of their phone contacts or social network are already using the application---or, indeed, to prompt the user to invite their friends to the application. Users are then able to quickly and easily initiate contact within the app.

However, this type of contact discovery is not well suited to Session's design nor its philosophy, as this functionality intrinsically requires users to link their accounts to a real-world identity to enable discovery. Further work should be done to investigate non-identifying ways to quickly bootstrap existing social networks into Session.

\subsection{Traffic Analysis}

The design of onion requests and Lokinet is to forward messages from their origin through to their final destination while traversing a number of routers (hops) with minimal delays. These types of onion routing networks are susceptible to traffic analysis attacks because of their reliance on the internet, a highly centralised system~\cites{Approximating_GPA_Tor,Recent_Tor_Attacks}. If both ends of an onion request connection between Alice and Bob are monitored by a co-opted ISP, then a path between the two can be discovered by inspecting Alice's outgoing encrypted packets, and correlating those with Bob’s incoming encrypted packets. Even if packets are padded to be a constant size, Alice’s ISP could introduce a delay (or drop packets) and, with the help of Bob’s ISP, watch as the delay or packet loss propagates into Bob’s connection.

\begin{figure}[htbp]
\vspace*{+0.1in}
\begin{center}
\begin{tabular}{ccc}
\includegraphics[width=16.5cm]{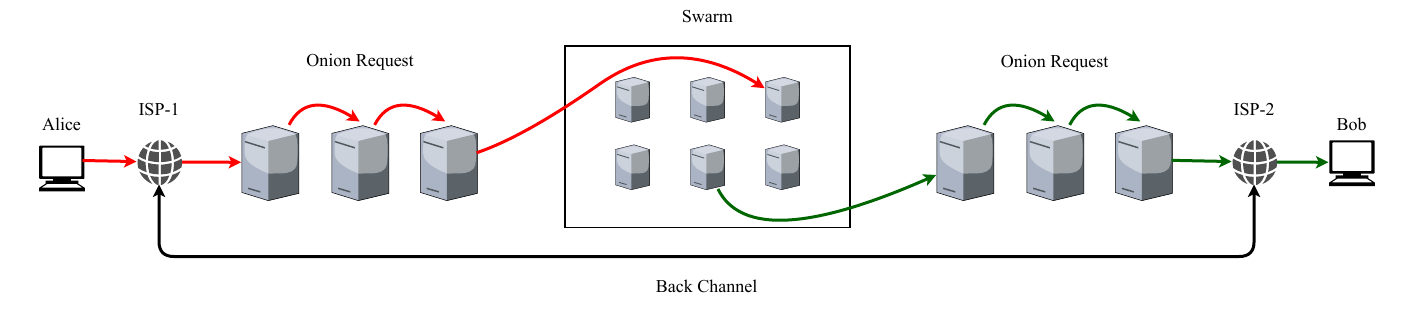}
\end{tabular}
\end{center}
\vspace*{-0.25in}
\caption{\textit{ISP 1 communicates with ISP 2 through a back channel. ISP 1 gives Alice's encrypted packets to ISP 2, who looks for packets with a similar structure being downloaded through Bob’s connection.}
\label{fig:backchannel}}
\end{figure}

Additionally, some state-level actors are known to directly monitor encrypted packets as data moves over the physical infrastructure of the internet without  requiring the explicit cooperation of ISPs~\cite{NSA_PRISM_Slides}.

\subsubsection{Traffic analysis solutions}

Session should investigate solutions as part of a future project to build a traffic analysis resistant layer for Session which operates on top of the existing Session Node network.

\textbf{Goals} 

Any solution should protect user privacy from global passive adversaries able to monitor all user-to-Session Node and Session Node-to-Session Node network traffic. It should also allow an adversary to own up to \(N-1\) of the set of nodes in a path without compromising the privacy of the path.

\textbf{Mix Networks (Mixnets)}

The most promising approach from initial research would be the deployment of an internal mixnet on top of the existing Session Node network. 

Mixnets are systems in which messages propagate through several layers of nodes (mixers) running on an overlay network. Each mixer holds and delays messages, usually until a certain number of messages have been sent or a set time interval is reached. The mixer then sends a batch of their held messages to another mixer, repeating the process. This makes traffic analysis difficult, as network adversaries are required to track batched messages. This has significant anonymity benefits, as even if only one mixer in a path is honest (not leaking each message's source and destination), malicious mixers (or network adversaries) are unable to effectively track the flow of messages. 

Local network adversaries, such as dishonest ISPs, are typically thwarted by padding messages and encrypting connections between hops. Timing attacks can be defeated using special decoy messages which loop back to the sender, acting as cover traffic.

On the other hand, it should also be considered that the delays introduced by mixnets negatively impact user experience.

\subsection{Community Discoverability}

Currently, community server operators may specify the location of their community with an IP address and public key combination or a Domain name and public key combination. In both cases, this produces an excessively long and non-human-readable string which is required to join the community. However, this is necessary to ensure non-reliance on the traditional certificate authority model for authentication and encryption.  

In the future, it should be possible for community operators to optionally buy SNS mappings to map human-readable names to IP/Domain and public key combinations; these mappings would be stored on the Arbitrum blockchain in a similar fashion to Session usernames in SNS. This would allow Session Communities to achieve a high level of domain security while maintaining human-readability. This scheme could be extended to allow communities to be discovered by downloading a list of registered community names from the Arbitrum blockchain via the Session Node network.

\subsection{Duress Codes}

Users may set a PIN or pattern lock to access the Session app, which adds additional security on top of any device-level passcodes. As an additional layer of security, users in the future should be able to specify a duress code which, if entered in lieu of the standard Session app PIN, will delete all client-side data from the device. This is useful in cases where users are forced to unlock their devices. 

\subsection{Remote Deletion}

Remote deletion would allow a user to specify the Account ID of a trusted contact and grant the power to remotely wipe Session data from their device. Once this contact was agreed upon, the trusted contact could generate a remote deletion message, if this message was received by the user’s device, it would initiate the immediate destruction of their local database and request the deletion of all messages stored by their swarm.

\section{Conclusion}

The internet is under ever-increasing surveillance by local, state, and corporate actors. With this mounting invasion of privacy, it is crucial that people have access to tools and applications that allow them to communicate online, while preserving the privacy of the contents and even the very existence of those conversations.

Leveraging privacy technologies including the Session Protocol, onion routing, and decentralised message storage, the Session application offers a secure messaging experience with minimal metadata leakage and robust encryption.

In addition to its fundamental privacy, Session is a user-friendly app which provides the functionalities expected from a modern messenger, including multi-device support, attachments, and group chats. Ongoing enhancements to Session are focused on introducing additional features to further strengthen security and privacy for its users, in alignment with their evolving needs.

\newpage

\section{Addendum}
\label{sec:Addendum}

\begin{itemize}
    \item The term \enquote{Account ID} will be replacing the currently used term \enquote{Session ID} in future versions of the application.
    \item The term \enquote{Recovery Password} will be replacing the currently used term \enquote{Recovery Phrase} in future versions of the application.
    \item The term \enquote{Session Node} will be replacing the currently used term \enquote{Service Node} in future versions of documentation. 
    \item The term \enquote{SNS} or \enquote{Session Name Service} will be replacing the currently used term \enquote{ONS} or \enquote{Oxen Name System}.
     \item The Session Token does not yet exist, instead Session Nodes currently earn rewards in a coin called Oxen which is issued on a layer 1 Cryptonote blockchain \cites{Oxen,CryptoNote}.
     \item The Session Node staking Reward Pool does not currently exist, currently Service Node rewards are received as Oxen coins which are minted in each newly created block.
    \item The Session Node list is currently called the Service Node list, and is held on the Oxen blockchain rather than the Arbitrum blockchain.
    \item The current implementation of passwords and PINs only allows Session Desktop users to set a password or PIN separate from their device password/PIN. More robust settings are coming to Session iOS and Android in the future. 
    \item The Groups system outlined in this paper, designed for fewer than 100 members, has not been incorporated into the live Session applications yet. Currently, Session employs a simpler groups approach \cite{Old_Session_Whitepaper}. Efforts are underway to integrate the enhanced protocol into the applications.
    \item Session employs a simpler method of ID blinding, which has minor deviations from the scheme presented in this paper.
    \item Message requests are currently supported only in one-on-one conversations; group invitations do not yet appear as message requests. 
\end{itemize}

\end{document}